\documentclass[preprint,superscriptaddress,floatfix]{revtex4-2}
\usepackage[T1]{fontenc}
\usepackage[dvipsnames]{xcolor}
\usepackage[colorlinks=true,allcolors=Blue,bookmarksopen=true]{hyperref}
\usepackage{graphicx}
\usepackage{siunitx}
\usepackage{stickstootext}
\usepackage[stix2]{newtxmath}
\usepackage{microtype}

\begin{document}
% \date{}
% Title
\title{Disentangling Core and Edge Mechanisms of the Density Limit in DIII-D Negative Triangularity Plasmas}

\author{R.~Hong}
\affiliation{University of California, Los Angeles, Los Angeles, California 90095, USA}
\author{P.H.~Diamond}
\affiliation{University of California, San Diego, La Jolla, California 92093, USA}
\author{O.~Sauter}
\affiliation{Ecole Polytechnique Fédérale de Lausanne (EPFL), Swiss Plasma Center (SPC), Lausanne, Switzerland}
\author{J.~Chen}
\affiliation{University of California, Los Angeles, Los Angeles, California 90095, USA}
\author{F.~Khabanov}
\affiliation{University of Wisconsin-Madison, Madison, Wisconsin 53706, USA}
\author{Z.~Li}
\affiliation{General Atomics, San Diego, California, 92121, USA}
\author{D.~Liu}
\affiliation{General Atomics, San Diego, California, 92121, USA}
\author{A.~Marinoni}
\affiliation{University of California, San Diego, La Jolla, California 92093, USA}
\author{G.R.~McKee}
\affiliation{University of Wisconsin-Madison, Madison, Wisconsin 53706, USA}
\author{T.L.~Rhodes}
\affiliation{University of California, Los Angeles, Los Angeles, California 90095, USA}
\author{F.~Scotti}
\affiliation{Lawrence Livermore National Laboratory, Livermore, California 94550, USA}
\author{K.E.~Thome}
\affiliation{General Atomics, San Diego, California, 92121, USA}
\author{G.R.~Tynan}
\affiliation{University of California, San Diego, La Jolla, California 92093, USA}
\author{M.A.~Van~Zeeland}
\affiliation{General Atomics, San Diego, California, 92121, USA}
\author{Z.~Yan}
\affiliation{University of Wisconsin-Madison, Madison, Wisconsin 53706, USA}
\author{L.~Zeng}
\affiliation{University of California, Los Angeles, Los Angeles, California 90095, USA}
\author{the DIII-D NT Team}
\noaffiliation{}

\begin{abstract}
    % \normalsize
    The density limit is investigated in the DIII-D negative triangularity (NT) plasmas which lack a standard H-mode edge. 
    We find the limit may not be a singular disruptive boundary but a multifaceted density saturation phenomenon governed by distinct core and edge transport mechanisms. 
    Sustained, non-disruptive operation is achieved at densities up to 1.8 times the Greenwald limit ($n_\mathrm{G}$) until the termination of auxiliary heating. 
    Systematic power scans show distinct power scalings for the core ($n_e \propto P_\mathrm{SOL}^{0.27\pm0.03}$) and edge ($n_e \propto P_\mathrm{SOL}^{0.42\pm0.04}$) density limits. 
    The edge density saturation is triggered by the onset of a non-disruptive, high-field side radiation front and the associated cooling, which clamps the edge density below $n_\mathrm{G}$. 
    In contrast, the core density continues to rise until it saturates, a state characterized by enhanced core turbulence. 
    Core transport evolves from a diffusive to an intermittent, avalanche-like state, as indicated by heavy-tailed probability density functions (kurtosis $\approx 6$), increased Hurst exponents, and a $1/f$-type power spectrum. 
    These findings suggest that the density limit in the low-confinement regime is determined by a combination of edge radiative cooling and core turbulent transport.
    This distinction provides separate targets for control strategies aimed at extending the operational space of future fusion devices.
\end{abstract}

\maketitle
% \clearpage

\section{Introduction}

High plasma density is a fundamental requirement for achieving a self-sustaining fusion reaction, as fusion power scales strongly with density \cite{lawsonCriteriaPowerProducing1957}.
However, tokamak operation is constrained by an operational density limit \cite{greenwaldNewLookDensity1988, greenwaldDensityLimitsToroidal2002, petriePlasmaDensityLimits1993}.
Exceeding this critical density leads to a rapid loss of confinement, often culminating in a plasma termination.
Overcoming this long-standing obstacle is therefore a critical step toward realizing economically viable fusion energy.

The conventional metric for the density limit is the empirical Greenwald scaling \cite{greenwaldNewLookDensity1988, greenwaldDensityLimitsToroidal2002}: $n_{\text{G}}\,(10^{20}\, \mathrm{m^{-3}})=\frac{I_\text{p}\,(\text{MA})}{\pi a^2 \,(\mathrm{m^2})}$, where $ I_\text{p} $ denotes the plasma current and $a$ the minor radius. 
This scaling is notably independent of heating power. 
This paradigm has been challenged by observations that the achievable density scales positively with heating power.
Recent theoretical modeling \cite{giacominFirstPrinciplesDensityLimit2022,singhZonalShearLayer2022,zancaPowerbalanceModelDensity2019} and database analyses \cite{manzPowerDependenceMaximum2023,huberImpactITERlikeWall2013,marisCorrelationLmodeDensity2025} suggest that the achievable density limit in the low-confinement (L-mode) regime may increase with heating power, potentially exceeding the conventional Greenwald density. 
Furthermore, the power-dependent density limits have also been observed in stellarators \cite{sudoScalingsEnergyConfinement1990,miyazawaDensityLimitStudy2008}, which lack plasma current, suggesting a more universal physical basis rooted in power balance and transport that transcends specific magnetic configurations.
Therefore, further dedicated investigations are needed to develop a physics-based understanding of the fundamental mechanisms governing density limits in magnetically confined plasmas.

The physics underpinning the L-mode density limit involves a complex interplay of several processes.
Turbulent transport \cite{rogersPhaseSpaceTokamak1998, labombardEvidenceElectromagneticFluid2005, hongEdgeShearFlows2018,longEnhancedParticleTransport2021, tokarSynergyAnomalousTransport2003}, radiative power losses \cite{lipschultzMarfeEdgePlasma1984,zancaPowerbalanceModelDensity2019, petriePlasmaDensityLimits1993}, and magnetohydrodynamic (MHD) instabilities \cite{suttropTearingModeFormation1997,gatesOriginTokamakDensity2012} have all been reported as key contributors. 
These processes do not act in isolation; instead, they are coupled \cite{greenwaldDensityLimitsToroidal2002} and ultimately constrain the plasma density.
Consequently, developing predictive models of density limits requires disentangling their individual roles and interdependencies.

To elucidate the fundamental physics of the L-mode density limit, it is advantageous to use a plasma configuration that approximates L-mode edge conditions over a wide range of operational parameters. 
This avoids the complexities introduced by the edge transport barriers (ETBs) and edge localized modes (ELMs), which are characteristic of conventional high-confinement (H-mode) plasmas. 
The negative triangularity (NT) plasmas \cite{kikuchiLmodeedgeNegativeTriangularity2019,thomeOverviewResults20232024}, characterized by an inverted “D”-shaped cross-section (see Fig.~\ref{fig:cross_section}), provides an ideal platform for this study. 
These plasmas exhibit improved core confinement \cite{austinAchievementReactorRelevantPerformance2019,marinoniDivertedNegativeTriangularity2021,nelsonRobustAvoidanceEdgeLocalized2023,codaEnhancedConfinementDiverted2022} while maintaining L-mode-like edge conditions.
While the NT edge is not identical to that of a conventional positive triangularity L-mode, its lack of H-mode ETBs and ELMs makes it a suitable proxy for isolating the core and edge transport phenomena that govern density limits without confounding factors.

This paper presents an examination of the physics governing the density limit in NT plasmas on the DIII-D tokamak. 
We report the achievement of sustained, non-disruptive high-density operation, where the plasma density can surpass the conventional Greenwald density by a factor of 1.8 at substantial input power. 
A systematic power scan reveals disparate scaling dependencies for the core ($n_{e} \propto P^{0.27\pm 0.03}_\text{SOL}$) and the separatrix ($n_{e} \propto P^{0.42\pm 0.04}_\text{SOL}$) regions. 
Detailed experimental measurements indicate that different mechanisms determine density limits in the core and edge plasmas.
At the plasma edge, the density is constrained by enhanced turbulence driven by radiative cooling.
In contrast, the core density is not directly limited by the radiative cooling but eventually saturates due to the emergence of enhanced, avalanche-like turbulence. 
These findings challenge the traditional paradigm of the density limit as a single boundary often associated with the empirical Greenwald scaling, advancing a more nuanced picture where distinct processes govern density saturation in different plasma regions.

In this work, the term `density limit' refers to a multifaceted density saturation phenomenon---the point at which the plasma density ceases to rise despite continued gas puffing. 
This distinguishes our focus from studies where the `limit' is defined by the onset of a disruption \cite{marisCorrelationLmodeDensity2025}.

The remainder of this paper is structured as follows. 
Section~\ref{sec:setup} describes the experimental arrangement, including the plasma configuration and the diagnostic suite employed in this study.
Section~\ref{sec:results} presents our experimental findings, organized into four subsections: 
(1) high-density operation beyond the Greenwald density and the power dependence of the core and edge density limits; 
(2) the temporal evolution of equilibrium profiles and turbulence approaching the density limit; 
(3) the edge density limit and the role of radiative edge cooling; 
(4) the core density limit and the dominant role of avalanche-like turbulence. 
Finally, Section~\ref{sec:conclusion} summarizes the key findings and
discusses their broader implications for fusion research.

\section{Experimental Arrangement}
\label{sec:setup}

\begin{figure}[tbp]
    \includegraphics[width=3.3in]{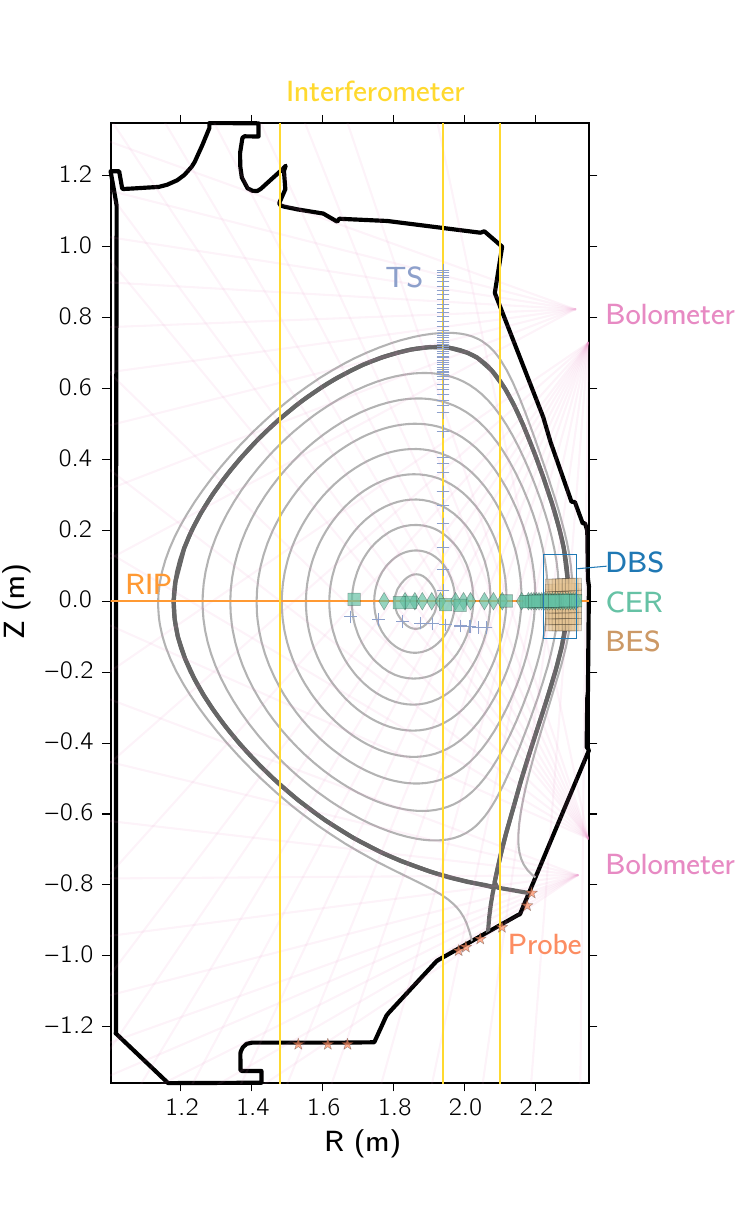}
    \caption{Cross-section of the DIII-D tokamak showing the negative triangularity plasma shape and the locations of key diagnostics used for this study.}
    \label{fig:cross_section}
\end{figure}

The experiment was conducted on the DIII-D tokamak \cite{luxonDesignRetrospectiveDIIID2002}, with auxiliary heating power provided by neutral beam injection (NBI) and varied between discharges. 
All discharges employed a lower single-null configuration with an open divertor geometry, and the ion $\mathbf{B} \times \nabla \mathbf{B}$ drift was directed away from the primary divertor, as illustrated in Fig.~\ref{fig:cross_section}. 
The plasma shape was characterized by negative triangularity at both the top and bottom ($\delta_\text{top} = -0.3$ and $\delta_\text{bot} = -0.62$), elongation ($\kappa \approx 1.3$), and an edge safety factor of $q_{95} = 4.1$.
This paper focuses specifically on discharges with toroidal field $B_t = 2.0$ T and plasma current $I_p = 0.6$ MA. 
Additional experiments exploring different plasma current and toroidal field combinations are reported in an accompanying paper \cite{sauterOperationGreenwaldDensity2025}.
In each discharge, plasma density was gradually increased via gas puffing from the top of the mainchamber. 
Consistent with the behavior of NT plasmas, no edge localized modes were observed in any discharge.

A suite of diagnostics monitored the evolving plasma state.
Figure~\ref{fig:cross_section} shows the locations of the key diagnostics.
Electron density and temperature in the main chamber were measured using the Thomson scattering (TS) system \cite{carlstromThomsonScatteringMeasurements2018}.
Charge exchange recombination (CER) spectroscopy \cite{chrystalImprovedEdgeCharge2016} measured the temperature, density, and toroidal and poloidal rotation velocities of carbon ions. 
These impurity ion profiles were used to compute radial electric field profiles ($ E_r $) using the carbon ion force balance equation. 
Bolometer arrays \cite{leonard2DTomographyBolometry1995} provided measurements of the radiated power distribution and total radiated power.

Several diagnostics were used to probe density fluctuations. 
A multichannel $\text{CO}_2$ interferometer \cite{vanzeelandFiberOpticTwocolor2006} measured the line-averaged density and its low-$k$ fluctuations along one horizontal midplane chord and three vertical chords at major radii of $R = 1.48\, \mathrm{m}$ (high-field side), $R = 1.94\, \text{m}$ (core region), and $R = 2.10\, \text{m}$ (low-field side). 
The radial interferometer polarimeter (RIP) \cite{chenFaradayeffectPolarimeterFast2018} also provided line-integrated measurements of electron density fluctuations at the mid-plane.
A 2D array of beam emission spectroscopy (BES) \cite{mckeeHighSensitivityBeam2006} measured ion-scale ($k_\theta<3\, \mathrm{cm^{-1}}$) density fluctuations at the edge. 
Two 8-channel Doppler back-scattering (DBS) systems \cite{peeblesNovelMultichannelCombfrequency2010, dambaEvaluationNewDIIID2022} measured turbulent flows and low-to-intermediate-$k$ ($2<k_\theta<10\,\mathrm{cm^{-1}}$) density fluctuations.
The radial locations and corresponding wavenumbers of the DBS measurements were calculated using the 3D ray-tracing code GENRAY
\cite{smirnovGeneralRayTracing1994}.

\section{Results}
\label{sec:results}

\subsection{Achieving High Density Beyond the Greenwald Limit}
\label{subsec:operation}

\subsubsection{Density Saturation Beyond the Greenwald Limit}

\begin{figure}[tbp]
    \includegraphics[width=3.3in]{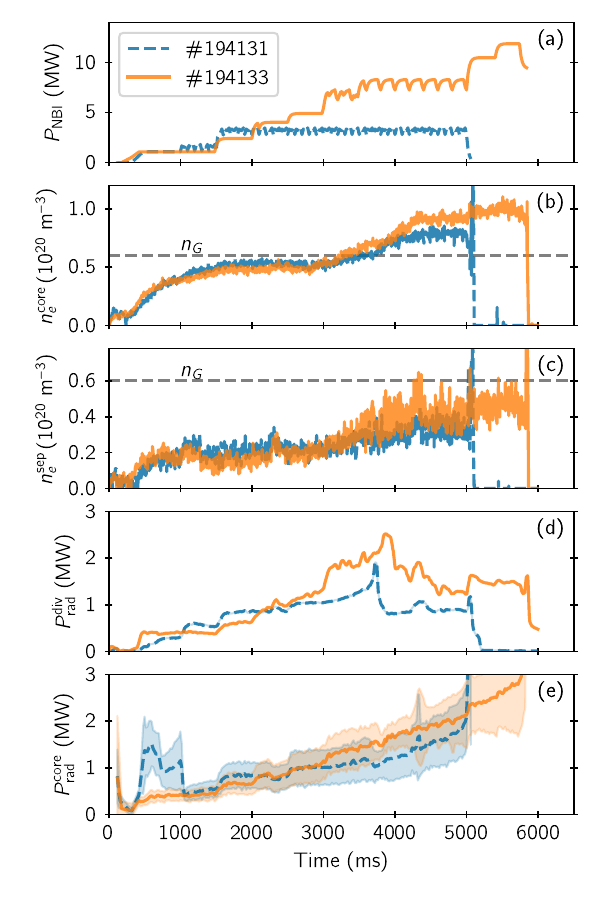}
    \caption{Time evolution of two representative NT discharges showing sustained, non-disruptive operation beyond the Greenwald density ($n_\mathrm{G}$, dashed line). Panels show: (a) NBI power, (b) core density ($\rho = 0.7$) exceeding $n_\mathrm{G}$, (c) separatrix density, (d) divertor radiated power, and (e) main chamber radiated power. Note the density saturation, rather than disruption, at high power.}
    \label{fig:time_history}
\end{figure}

This study shows that the density limit in these plasmas can surpass the conventional Greenwald limit and manifests as a saturation phenomenon.
This is shown in Fig.~\ref{fig:time_history}, which presents the time evolution of key parameters in two representative discharges with different levels of auxiliary heating from NBI.
Across the experiment, the NBI power was varied from 3 MW up to 13 MW.
As shown in Figs.~\ref{fig:time_history}(b) and \ref{fig:time_history}(c), both core and separatrix densities increase with input power.
The core density at $ \rho\approx0.7 $, which serves as a proxy for the line-averaged density, exceeds the Greenwald density ($n_\text{G}$).
The discharges with medium to high NBI power ($P_\mathrm{NBI} > 4~\mathrm{MW}$) transitioned into a phase of density saturation that persisted until the termination of auxiliary heating.

The radiated power in the divertor [Fig.~\ref{fig:time_history}(d)] and the main chamber [Fig.~\ref{fig:time_history}(e)] also increase with density, indicating enhanced radiative losses.
The spikes in the divertor radiation power [Fig.~\ref{fig:time_history}(d)] mark the development of a radiation front that propagates from the divertor towards the high-field side (HFS).

The line-averaged density, inferred from Thomson scattering profiles, reached $1.8 \times n_\text{G}$ in the high-power discharge. 
For comparison, the CO$_2$ interferometer measured a line-averaged density of $2.0\times n_\text{G}$ \cite{thomeOverviewResults20232024,sauterOperationGreenwaldDensity2025}. 
This discrepancy arises because the interferometer's measurement includes the substantial scrape-off layer density present during high-density operation. 
Therefore, Thomson scattering measurements are used for the analysis in this study.

\subsubsection{Disparate Power Scaling for Core and Edge Density Limit}

Having established that the density limit in these NT plasmas is a saturation phenomenon, we investigate its dependence on heating power. 
The analysis correlates the maximum achieved density with the power flowing into the scrape-off layer ($P_\mathrm{SOL}$), defined as $P_\mathrm{SOL} = P_\mathrm{in} - P_\mathrm{rad,core} - \mathrm{d}W/\mathrm{d}t$, accounting for total input power, core radiated power, and changes in stored energy. 
This parameter represents the effective supplied heating. 
All values are averaged over a 100 ms window at the point of maximum density.

\begin{figure}[tbp]
    \includegraphics[width=3.5in]{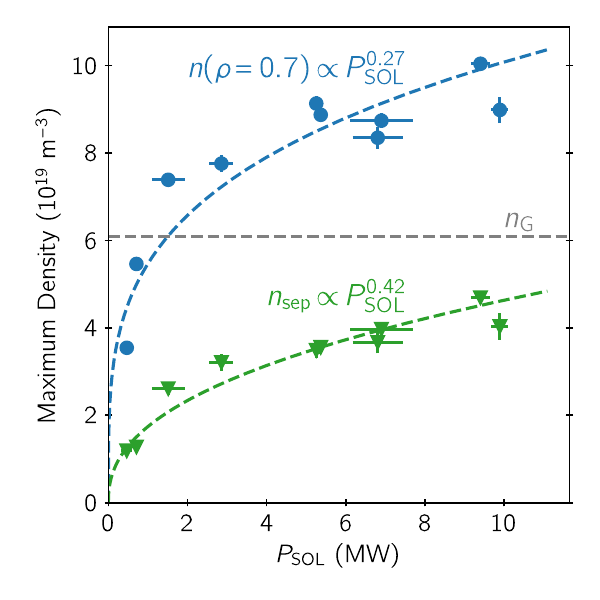}
    \caption{Disparate power scalings for the core and edge density limits. The maximum achieved density is plotted as a function of power into the scrape-off layer ($P_\mathrm{SOL}$) for the core ($\rho=0.7$) and the separatrix. The core density exhibits a weak scaling ($\propto P_\mathrm{SOL}^{0.27}$), while the edge density shows a stronger dependence ($\propto P_\mathrm{SOL}^{0.42}$), implying distinct limiting mechanisms. The error bars represent the standard error of the mean calculated over a 100 ms window.}
    \label{fig:power_dependence}
\end{figure}

A systematic power scan reveals that the core and edge densities respond differently to $P_\mathrm{SOL}$, as shown in Fig.~\ref{fig:power_dependence}. 
The maximum core density (at $\rho = 0.7$) exhibits a power scaling of $n_{e,\text{core}} \propto P_\mathrm{SOL}^{0.27\pm0.03}$. 
In contrast, the separatrix density displays a stronger power dependence, scaling as $n_{e,\text{sep}} \propto P_\mathrm{SOL}^{0.42\pm0.04}$. 
This is consistent with edge density limit scalings reported on other tokamaks; for example, it is in agreement with the scaling of $n_\mathrm{sep} \propto P_\mathrm{SOL}^{0.38\pm 0.08}$ recently reported by ASDEX Upgrade \cite{manzPowerDependenceMaximum2023}. 
This relatively strong power dependence of the edge density limit is also reminiscent of the density limit observed in stellarators, where power balance is known to be the dominant constraint~\cite{sudoScalingsEnergyConfinement1990,miyazawaDensityLimitStudy2008}.

These different power scalings indicate that the density saturation is not governed by a single, global mechanism. 
Instead, it suggests that distinct physical processes, with different sensitivities to heating power, are responsible for limiting the density in the core and edge regions.

\subsection{Profile Evolution and Confinement Degradation Towards Density Limit}
\label{subsec:profile_evolution}

\subsubsection{Reduced Temperature and Rotational Shear at High Density}

\begin{figure}[tbp]
    \includegraphics[width=3.7in]{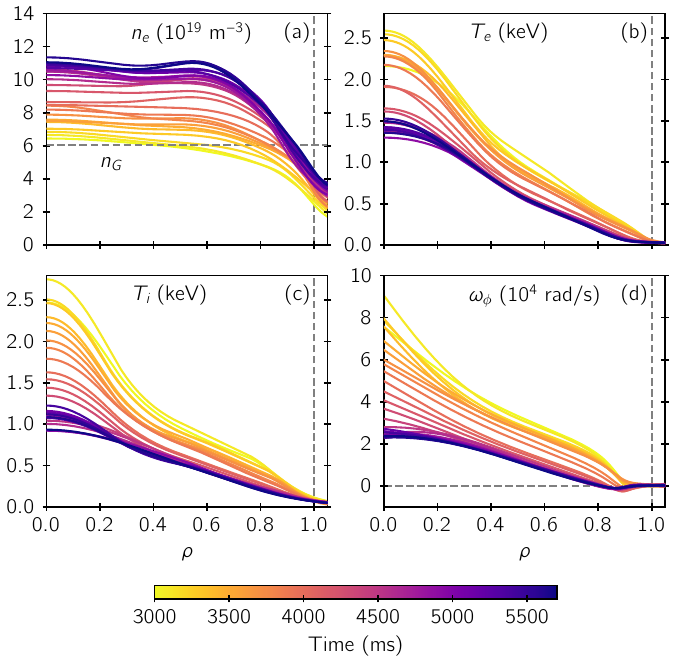}
    \caption{Progressive profiles approaching the density limit. Temporal evolution of (a) electron density, (b) electron temperature, (c) ion temperature, and (d) toroidal rotation. As density rises (yellow to purple), the temperature and rotation profiles collapse, indicating a loss of thermal and momentum confinement. The profiles are generated from cubic spline fits to data averaged over a 40 ms time window.}
    \label{fig:profiles}
\end{figure}

As the plasma approaches the density limit, its equilibrium profiles evolve with degradation of thermal confinement.
This evolution, shown for a representative high-power discharge in Fig.~\ref{fig:profiles}, provides further insight into the underlying transport dynamics.

The electron density profile [Fig.~\ref{fig:profiles}(a)] rises across the radius but remains flat in the core, indicating persistent particle transport at high densities.
The degraded confinement is accompanied by a loss of thermal energy, as both the electron and ion temperature profiles collapse [Figs.~\ref{fig:profiles}(b), \ref{fig:profiles}(c)].
The edge $T_e$ profile flattens significantly, a characteristic signature of intensified radiative cooling at the plasma boundary.

\begin{figure}[tbp]
    \includegraphics[width=3.6in]{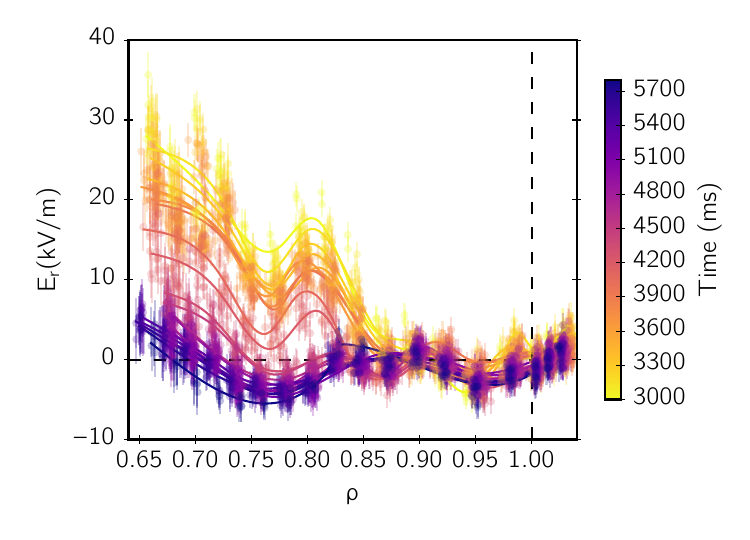}
    \caption{The evolution of the radial electric field ($E_r$) approaching the density limit. As the discharge evolves (yellow to purple), the core $E_r$ shear, a key mechanism for turbulence suppression, progressively weakens and ultimately vanishes.}
    \label{fig:Er}
\end{figure}

Another critical change occurs in the plasma rotation. 
The toroidal rotation profile [Fig.~\ref{fig:profiles}(d)], initially peaked in the core, progressively collapses with increasing density. 
This loss of core rotation eliminates the dominant component of the radial electric field ($E_r$), reducing turbulence suppression of flow shear decorrelation. 
As shown in Fig.~\ref{fig:Er}, the core $E_r$ shear, initially strong, diminishes and ultimately vanishes as the density limit is approached.

This sequence of events, from intensified edge cooling to the collapse of core rotation and $E_r$ shear, indicates that the plasma's ability to suppress turbulence deteriorates with rising density.
The loss of this stabilizing flow shear enables enhanced turbulent transport that governs the core density limit, as subsequent sections will demonstrate.

\subsubsection{Divergent Evolution of Turbulence in Different Regions}

\begin{figure}[tbp]
    \includegraphics[width=4in]{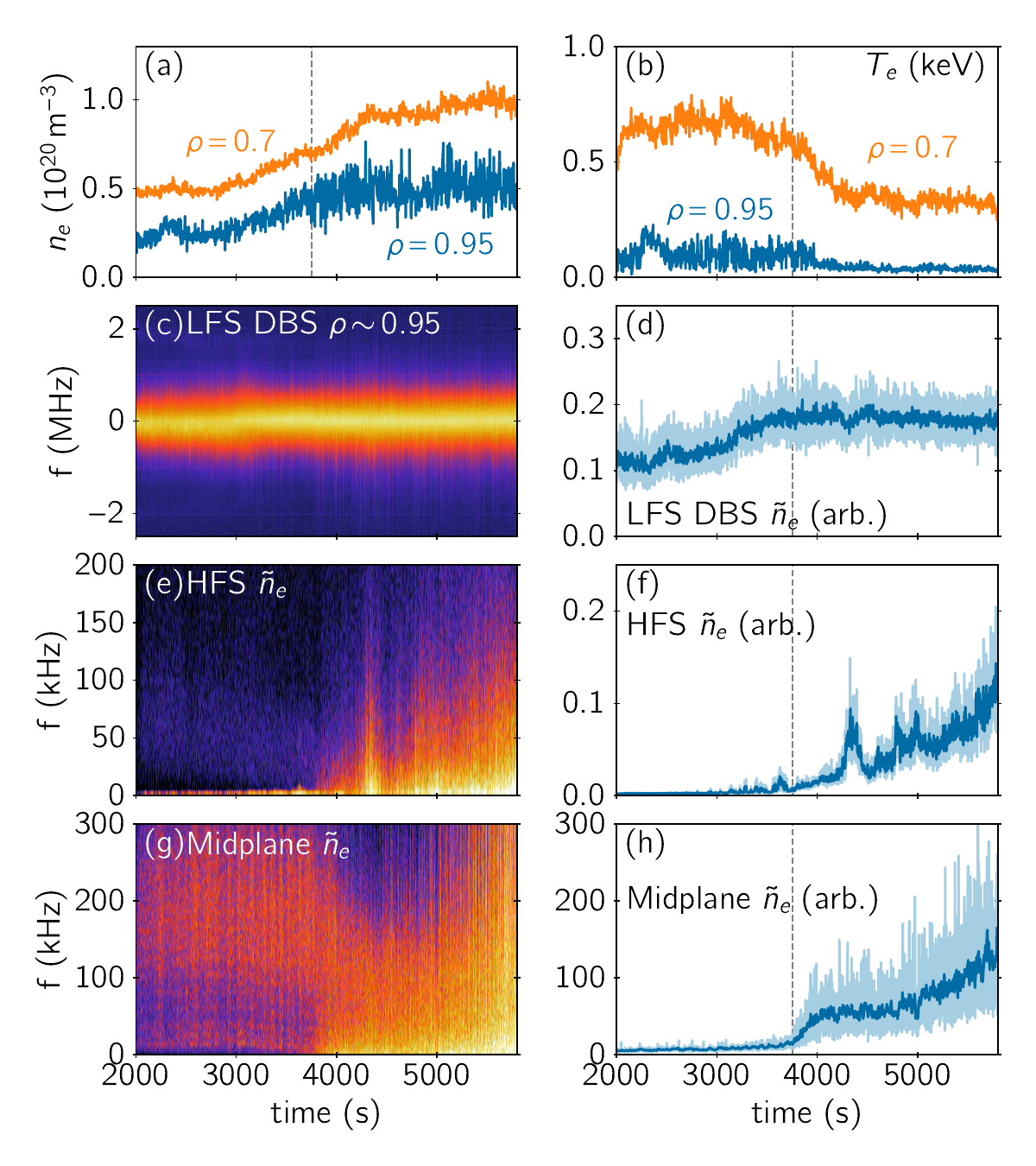}
    \caption{Divergent evolution of turbulent fluctuations (shot 194133). A critical transition occurs at $\sim$3750 ms (dashed line). (d) Localized LFS turbulence saturates after the transition, while line-integrated fluctuations on the (f) HFS and (h) midplane intensify, revealing a strong poloidal asymmetry. Panels (a) and (b) show density and $T_e$, while (c), (e), and (g) are the corresponding spectrograms.}
    \label{fig:turbulence_time}
\end{figure}

The degradation of thermal confinement described above coincides with divergence in turbulent activity across the plasma, as illustrated in Fig.~\ref{fig:turbulence_time} for a high-power discharge. 
A critical transition occurs at approximately 3750~ms (vertical dashed line), marking the onset of strong edge cooling. 
Before this event, the turbulence evolution is relatively uniform; afterward, it diverges sharply between different spatial regions.

Initially, both core and edge densities rise together [Fig.~\ref{fig:turbulence_time}(a)], accompanied by a steady increase in the amplitude of local, intermediate-$k$ turbulence at the low-field side (LFS) midplane ($\rho \approx 0.95$), as measured by Doppler backscattering [Fig.~\ref{fig:turbulence_time}(d)]. 
This period of evolution ends abruptly, initiated by a transition at approximately 3750 ms.
Immediately following this trigger, the LFS turbulence saturates. 
A clamping of the edge density and a sharp drop in local electron temperature also start, which are fully established by $t=4000$~ms [Fig.~\ref{fig:turbulence_time}(b)].

In contrast, turbulence in other regions intensifies. 
Line-integrated, low-$k$ density fluctuations on the high-field side (HFS), measured by the CO$_2$ interferometer, increase substantially [Figs.~\ref{fig:turbulence_time}(e), \ref{fig:turbulence_time}(f)]. 
A similar growth is observed in the line-integrated midplane fluctuations measured by the RIP diagnostic [Figs.~\ref{fig:turbulence_time}(g), \ref{fig:turbulence_time}(h)].

This divergent evolution---saturation of localized turbulence on the LFS versus continued, strong growth of HFS and core fluctuations---indicates a strong poloidal asymmetry in the transport dynamics.
This asymmetry indicates that the density limit is not a uniform process but is likely triggered by a localized mode on the high-field side—likely the radiative phenomenon that we investigate next.

\subsection{The Edge Density Saturation and the Role of the HFS Radiation Front}

\subsubsection{Visualizing the HFS Radiation Front}

\begin{figure}[tbp]
    \includegraphics[width=4in]{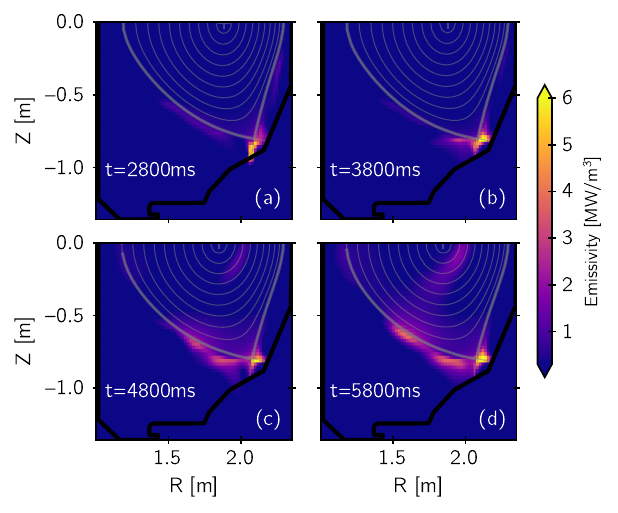}
    \caption{Time-sequence of reconstructed bolometry measurements (shot 194133), showing the formation and propagation of the HFS radiation front. The front begins near the X-point (b), detaches from the divertor (c), and moves upward along the HFS separatrix (d).}
    \label{fig:bolometry}
\end{figure}

The poloidally asymmetric turbulence described previously points to a localized HFS mode as the trigger for the edge density saturation. 
Here, we examine the 2D radiated power distribution, reconstructed from bolometer measurements, to visualize the formation and evolution of this mode. 
Figure~\ref{fig:bolometry} presents a time-sequence of this evolution. 
While geometrically similar to a disruptive MARFE~\cite{lipschultzMarfeEdgePlasma1984,lipschultzReviewMARFEPhenomena1987}, the feature observed here is a quasi-stable, non-disruptive radiation front that is directly linked to the observed density saturation.
Its onset is sometimes referred to as a radiative mode to emphasize its role in triggering the transition to a high-transport state, though it evolves into a persistent radiative structure reminiscent of the High-Field-Side High-Density (HFSHD) phenomenon observed in other devices \cite{potzelFormationHighDensity2015}.

The evolution of this front proceeds in stages.
Early in the discharge ($ t=2800 $ ms), before the density ramp, the radiation is concentrated at the inner divertor target, characteristic of a standard attached divertor [Fig.~\ref{fig:bolometry}(a)].
As plasma density increases ($t=3800$ ms), an intense radiation zone emerges near the X-point, signaling the birth of the radiative front [Fig.~\ref{fig:bolometry}(b)]. 
This event coincides with the onset of the divergent turbulence seen in Fig.~\ref{fig:turbulence_time}.
With a further increase in density ($t=4800$ ms), this front detaches from the divertor and propagates upward along the HFS separatrix [Fig.~\ref{fig:bolometry}(c)]. 
This movement is strongly correlated with the localized edge cooling and turbulence saturation.
Finally ($t = 5800$ ms), the front moves further inboard and its emissivity intensifies [Fig.~\ref{fig:bolometry}(d)].

This sequence provides evidence of a developing radiation front on the HFS. 
While the specific radial location of the peak emissivity has some uncertainty inherent in the tomographic reconstruction, the overall structure and HFS localization of the radiation front are robust features of the data.
The role of this dynamic radiative structure will be discussed in detail later, as it appears to be a key triggering event, coinciding with the saturation of the edge density.
Notably, the eventual discharge termination was caused by a drop in NBI power that triggered an $n=1$ locked mode and was not a direct consequence of the radiation front.

\subsubsection{Radiative Cooling and the Onset of HFS Turbulence}

\begin{figure}[tbp]
    \includegraphics[width=3.5in]{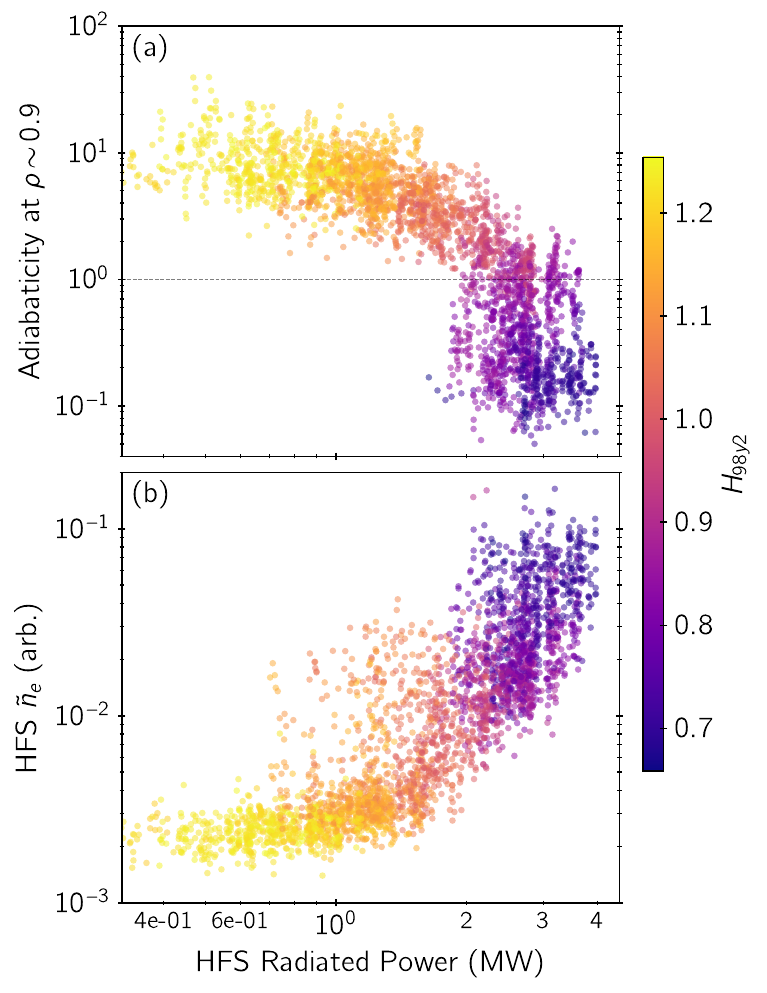}
    \caption{Onset of a non-adiabatic, turbulent edge state at HFS radiated power. Above a critical threshold of $\sim 2$~MW, (a) the edge adiabaticity parameter drops below unity, and (b) the HFS density fluctuation amplitude surges. The color scale ($H_{98y2}$) shows this coincides with a transition from high (yellow) to low (purple) confinement in time sequence.}
    \label{fig:radiative_turb_scaling}
\end{figure}

The formation of the HFS radiation front actively cools the plasma and triggers a significant shift in the local turbulent state. 
Figure~\ref{fig:radiative_turb_scaling} provides a quantitative analysis of this transition, using data compiled from multiple discharges.
As the HFS radiated power increases beyond a critical threshold, the edge plasma parameters and HFS turbulence changes significantly.
The RMS amplitude of the line-integrated HFS density fluctuations ($\tilde{n}_\mathrm{HFS}$), measured by the CO$_2$ interferometer passing through the radiation front, serves as a proxy for the mode's strength.

The effect of edge cooling can be characterized by the dimensionless adiabaticity parameter \cite{wakataniCollisionalDriftWave1984}, $\alpha_\mathrm{adia}=k^2_\parallel v^2_{te}/ \omega \nu_{ei} \propto \frac{T_e^{5/2}}{n_e}$, which represents the comparison between parallel electron motion and turbulent timescales. 
Here, $k_\parallel$ is the parallel wavenumber of turbulence, $v_{te}$ is the electron thermal velocity, $\omega$ is the turbulence frequency which can be calculated using the power-weighted mean frequency of the underlying fluctuations, and $\nu_{ei}$ is the electron-ion collision frequency.
When $\alpha_\mathrm{adia} < 1$, the plasma enters a non-adiabatic regime where drift-wave instabilities are expected to grow. 
As shown in Fig.~\ref{fig:radiative_turb_scaling}(a), the onset of the strong HFS radiation drives $\alpha_\mathrm{adia}$ to drop sharply below 1.

This transition into a non-adiabatic state coincides with a surge in HFS turbulence. 
Figure~\ref{fig:radiative_turb_scaling}(b) shows that the RMS amplitude of HFS density fluctuations ($\tilde{n}_e$) increases sharply as the radiated power crosses the same threshold. 
The entire event is correlated with a degradation in plasma performance, as indicated by the drop in the $H_{98y2}$ confinement factor (color scale) in time sequence. 
This finding indicates that as the HFS radiation front intensifies, it alters the local plasma state by driving it into a non-adiabatic regime, which is correlated with a surge in turbulence that may ultimately saturate the edge density.

\subsubsection{Clamping of the Edge Density}

The preceding observations suggest that the HFS radiation front can be a key trigger for the edge density saturation.
To test this hypothesis quantitatively, we correlate the local edge density with the mode's strength, again using the RMS amplitude of the line-integrated HFS density fluctuations ($\tilde{n}_\mathrm{HFS}$) as a proxy.

\begin{figure}[tbp]
    \includegraphics[width=3.5in]{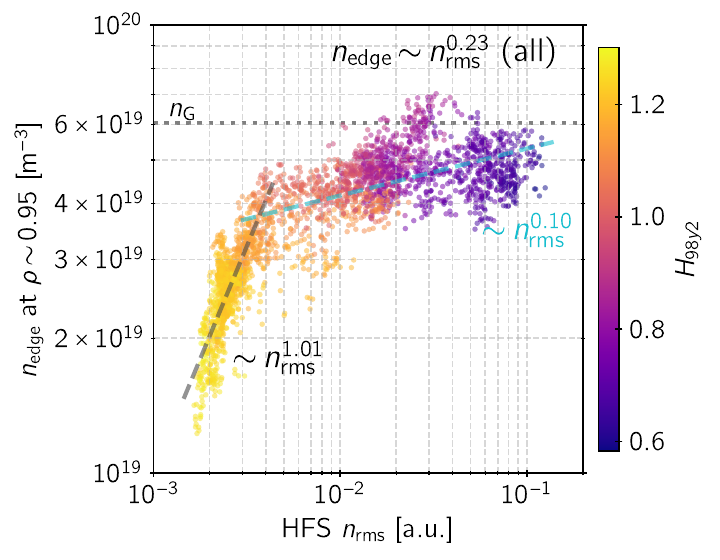}
    \caption{Bifurcation in edge transport, showing the clamping of edge density by HFS turbulence. At low turbulence, density and fluctuations grow together (yellow points). Above a critical threshold, the system transitions to a degraded state (purple points) where the edge density saturates completely despite a large increase in turbulence amplitude.}
    \label{fig:edge_turb_scaling}
\end{figure}

Figure~\ref{fig:edge_turb_scaling} plots the edge density (at $\rho = 0.95$) against $\tilde{n}_\mathrm{HFS}$ on a log-log scale. The data, compiled from multiple discharges, shows a bifurcation in the plasma behavior, defined by two distinct operational regimes that are distinguished by the energy confinement factor, $H_{98y2}$.

At low turbulence amplitudes, the plasma is in a high-confinement state (high $H_{98y2}$, yellow points). 
In this phase, the edge density and turbulence grow together, following a nearly linear relationship ($n_{e,\mathrm{edge}} \propto \tilde{n}_\mathrm{HFS}^{1.0}$). 
This represents a state where turbulence-driven transport has not yet begun to limit the density.

This linear relationship breaks down abruptly once the turbulence level surpasses a critical threshold, at which point the system transitions to a degraded-confinement state (low $H_{98y2}$, purple points). 
While the turbulence amplitude continues to grow by nearly an order of magnitude, the edge density saturates, with its scaling flattening significantly ($n_{e,\mathrm{edge}} \propto \tilde{n}_\mathrm{HFS}^{0.1}$).

This bifurcation suggests that the edge density limit is not a gradual process, but a sharp transition associated with the onset of a radiative mode.
Once initiated, the mode clamps the edge density at a value consistently below the Greenwald limit, leading to a significant increase in HFS turbulence. 
The apparent overall relationship across both regimes ($n_{e,\mathrm{edge}} \propto \tilde{n}_\mathrm{HFS}^{0.23}$) also suggests the sharpness of this critical transition.

\subsubsection{Breakdown of Zonal Flow Regulation}

\begin{figure}[tbp]
    \includegraphics[width=4.5in]{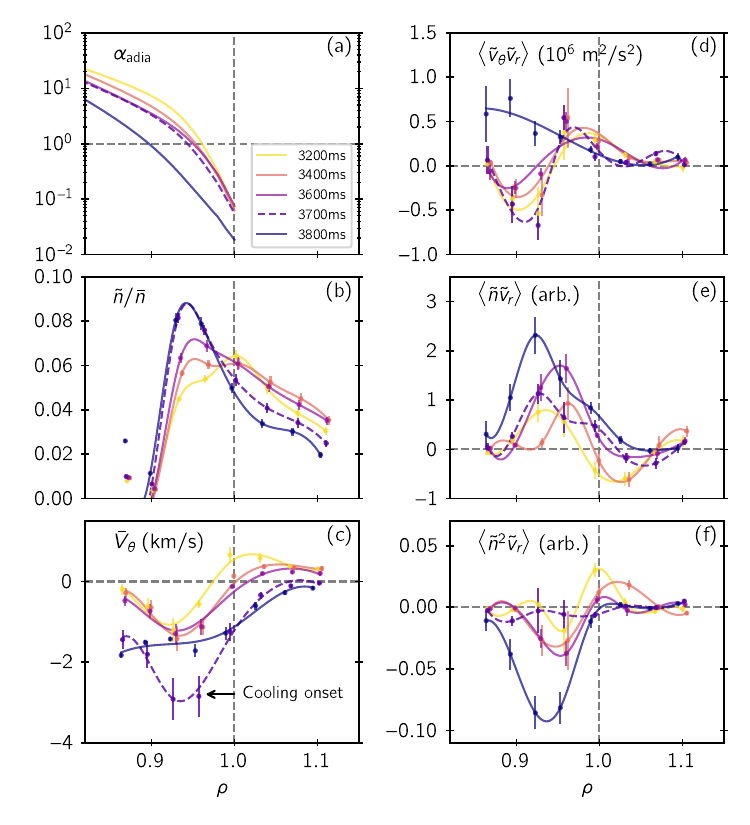}
    \caption{Radial profiles of LFS edge turbulence characteristics during radiative cooling in a low-power NT discharge (shot 194131): (a) adiabaticity parameter, (b) normalized density fluctuation amplitude, (c) mean poloidal phase velocity, and (d) Reynolds stress \(\langle \tilde{v}_\theta \tilde{v}_r \rangle\), (e) particle flux \(\langle \tilde{n} \tilde{v}_r \rangle\), and (f) turbulent spreading flux \(\langle \tilde{n}^2 \tilde{v}_r \rangle\). The profiles are colored by the time from lighter to darker. Particularly, the profiles corresponding to the onset of the radiative cooling (about 3700 ms) are marked by dashed curves in purple.}
    \label{fig:turbulence_vel}
\end{figure}

The onset of the HFS radiation front has global consequences, leading to a failure of turbulence self-regulation across the plasma edge.
We use BES measurements on the low-field side to observe this breakdown in detail (Fig.~\ref{fig:turbulence_vel}), revealing the microphysical mechanism responsible for enhanced transport.
It is worth mentioning that BES measurements are taken from a different discharge (shot 194131) with lower NBI power, which results in a slightly earlier onset of the radiative cooling.

The trigger for this breakdown is the transition to the non-adiabatic regime.
As the edge cools with the onset of the radiation front ($t \approx 3700$~ms), the dimensionless adiabaticity parameter, $\alpha_\mathrm{adia}$, drops below unity [Fig.~\ref{fig:turbulence_vel}(a)]. 
This transition is expected to alter drift-wave turbulence, weakening the mechanisms that generate stabilizing zonal flows~\cite{wakataniCollisionalDriftWave1984,hongEdgeShearFlows2018}.

While the fluctuation amplitude on the LFS saturates (as seen in Fig.~\ref{fig:turbulence_time}), the underlying dynamics show a flattening of the Reynolds stress profile [$\langle \tilde{v}_\theta \tilde{v}_r \rangle$, Fig.~\ref{fig:turbulence_vel}(d)]. 
This collapse of the primary nonlinear drive for zonal flows (i.e., the Reynolds force, \(\mathcal{F}_\text{Re} = - \partial_r \langle \tilde{v}_\theta \tilde{v}_r \rangle\)) signifies a breakdown in the zonal flow regulation mechanism.
Concurrently, the mean poloidal phase velocity of turbulence [$\langle \tilde{v}_\theta \rangle$, Fig.~\ref{fig:turbulence_vel}(c)] also collapses.

This breakdown in zonal flow regulation allows for increased particle loss. 
The outward turbulent particle flux [$\langle \tilde{n} \tilde{v}_r \rangle$, Fig.~\ref{fig:turbulence_vel}(e)] increases significantly, while the turbulent spreading flux [$\langle \tilde{n}^2 \tilde{v}_r \rangle$, Fig.~\ref{fig:turbulence_vel}(f)] indicates that turbulence intensity is propagating inward from the separatrix. 
These measurements suggest a sequence: the formation of the HFS radiation front triggers edge cooling, which pushes the entire edge into a non-adiabatic state. 
This disables turbulence self-regulation and hence the zonal flow production on the LFS, enabling enhanced particle transport and contributing to the clamping of the edge density profile.

\subsection{The Core Density Saturation and the Emergence of Avalanche-like Turbulence}

While the edge density saturates below the Greenwald limit, the core density continues to rise but eventually plateaus at a value significantly above $ n_\text{G} $.
This decoupling implies that a separate, core-localized mechanism must govern the core density profile.
Here, we show that this mechanism is not a simple enhancement of diffusive transport, but is correlated with a transition in its character to intermittent, avalanche-like behavior.
This transition can be enabled by the collapse of the stabilizing $E_r \times B$ shear, as shown previously in Fig.~\ref{fig:Er}.

\subsubsection{Statistical Signatures of Avalanche-like Transport}

The statistical properties of the density fluctuations, measured by the RIP diagnostic at the midplane, reveal three distinct signatures of this transition. 

\begin{figure}[tbp]
    \includegraphics[width=3.5in]{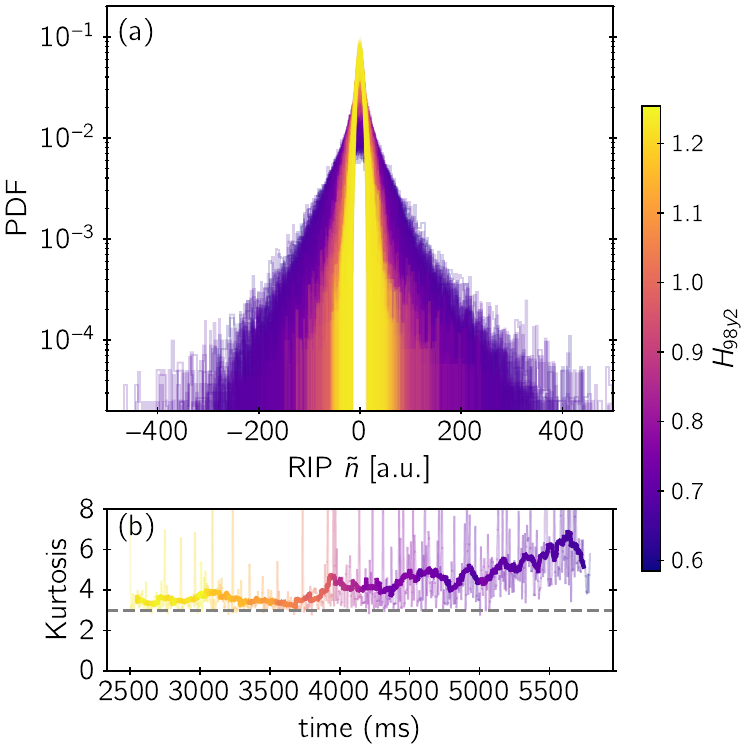}
    \caption{Statistical signatures of the transition to intermittent, avalanche-like core transport. (a) The probability density function (PDF) of core fluctuations evolves from Gaussian (yellow) to heavy-tailed (purple). (b) This growing intermittency is quantified by the kurtosis, which rises far above the Gaussian value of 3.}
    \label{fig:rip_pdf_kurtosis}
\end{figure}

First, the probability density function (PDF) of the fluctuation amplitudes evolves from Gaussian to heavy-tailed. 
Early in the discharge the PDF is Gaussian [yellow in Fig.~\ref{fig:rip_pdf_kurtosis}(a)], characteristic of diffusive transport. 
As the density limit is approached, the PDF develops non-Gaussian tails [purple in Fig.~\ref{fig:rip_pdf_kurtosis}(a)]. 
This signifies a shift to a transport regime dominated by large-amplitude, intermittent events. 
The growing intermittency is quantified by the kurtosis of the PDF, which rises from the Gaussian value of 3 to about 6 [Fig.~\ref{fig:rip_pdf_kurtosis}(b)].

\begin{figure}[tbp]
    \includegraphics[width=3.5in]{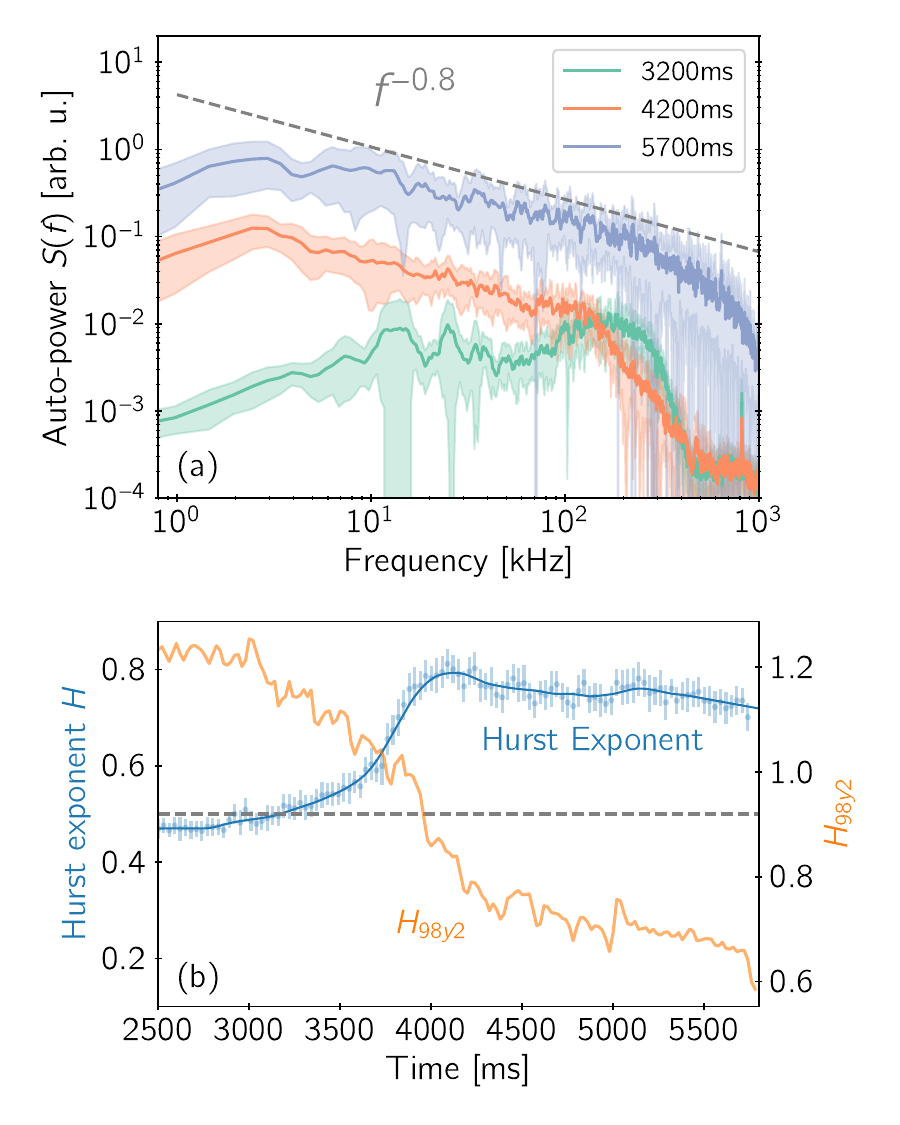}
    \caption{Evidence of self-organized criticality and long-range correlations in core turbulence. (a) The power spectrum develops a $1/f$-type scaling characteristic of avalanche dynamics. (b) The Hurst exponent ($H$) rises from $\sim 0.5$ (diffusive) to $\sim 0.8$ (persistent, correlated), coinciding with a drop in confinement ($H_{98y2}$).}
    \label{fig:rip_autopower}
\end{figure}

Second, the frequency spectrum of the turbulence develops a $1/f$-type power-law scaling. 
As the plasma evolves with rising density, a distinctive power-law scaling emerges at a wide range of frequencies [Fig.~\ref{fig:rip_autopower}(a)]. 
Such a spectrum is a classic signature of self-organized criticality (SOC) and avalanche dynamics, where transport occurs via cascading events spanning a wide range of scales~\cite{hahmMesoscopicTransportEvents2018, politzerObservationAvalanchelikePhenomena2000, diamondDynamicsTurbulentTransport1995}.

Third, the transport acquires long-range temporal correlations, quantified by the Hurst exponent ($H$), a measure of the persistence in a time series [Fig.~\ref{fig:rip_autopower}(b)]. 
The Hurst exponent rises from $H \approx 0.5$ (random-walk events, diffusive transport) to $H \approx 0.8$ as the density limit is approached. 
A value of $H > 0.5$ indicates persistent, correlated dynamics, another key feature of avalanche-like transport. 
This increase in $H$ correlates with a substantial decrease in the energy confinement factor $H_{98y2}$, linking the emergence of this non-diffusive transport regime to the degradation in plasma performance.

Collectively, these three independent signatures---heavy-tailed PDFs, a $1/f$ power spectrum, and an elevated Hurst exponent---are consistent with a scenario in which turbulent transport undergoes a transition from a diffusive to an intermittent, avalanche-like regime, which may be the mechanism for the core density limit.

It is worth noting that the RIP diagnostic is line-integrated, and its signal inherently contains contributions from the entire plasma profile along its midplane chord, including both the core and the edge.
However, local measurements of the edge turbulence on the low-field side midplane from both DBS and BES do not exhibit the signatures of avalanche-like transport that we observe in the RIP data.
Therefore, while the RIP signal contains contributions from the full profile, its change in character towards an avalanche-like state can serve as an indicator of the changing dynamics of the core transport.

\subsubsection{Core Density Saturation by Avalanche-like Turbulence}

The emergence of avalanche-like dynamics provides a hypothesis for the core density limit: the large, intermittent transport events ultimately clamp the density profile. 
We test this hypothesis by correlating the core density (at $\rho \approx 0.7$) with the RMS amplitude of the midplane turbulence measured by RIP.

\begin{figure}[tbp]
    \includegraphics[width=3.5in]{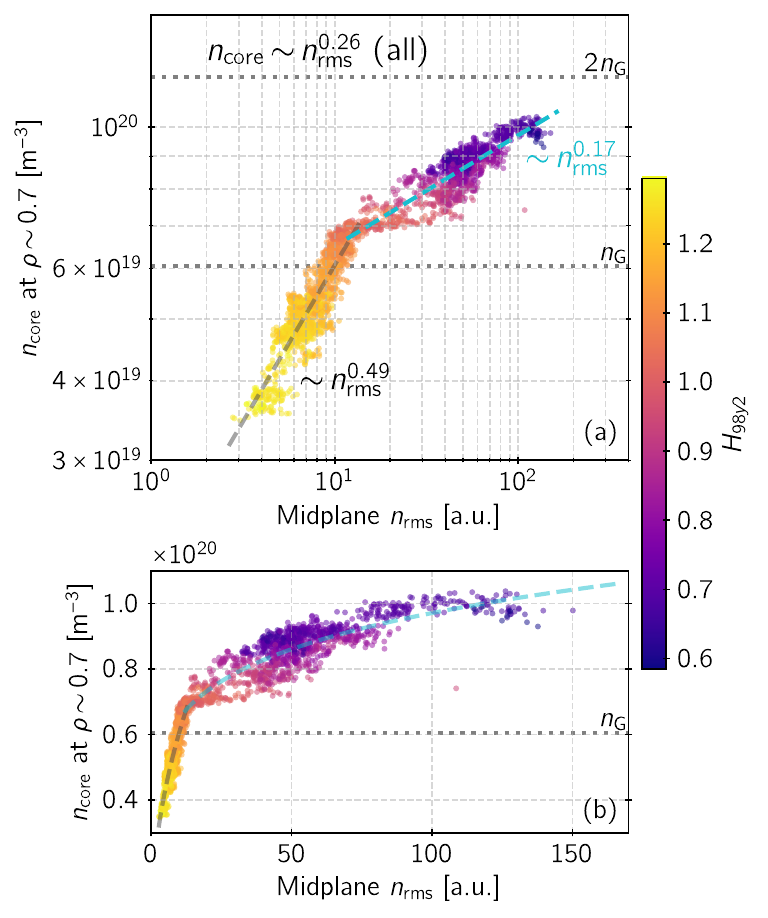}
    \caption{Saturation of core density due to enhanced turbulent transport, plotted in (a) log scale and (b) linear scale. The relationship between core density and turbulence bifurcates into two regimes. In the high-confinement state (yellow), density rises with turbulence ($\tilde{n}_\mathrm{rms}^{0.49}$). In the avalanche-like state (purple), the scaling flattens significantly ($\tilde{n}_\mathrm{rms}^{0.17}$).}
    \label{fig:core_turb_scaling}
\end{figure}

As shown in Fig.~\ref{fig:core_turb_scaling}, data compiled from multiple discharges reveal a clear bifurcation between two transport regimes, distinguished by their confinement properties ($H_{98y2}$).

In the low-turbulence, high-confinement regime ($H_{98y2} > 1.0$, yellow points), which corresponds to the diffusive state, the core density rises with increasing turbulence amplitude, following a power-law scaling of $n_{e,\mathrm{core}} \propto \tilde{n}_\mathrm{rms}^{0.49}$. 
This phase represents a state where fueling remains effective despite the presence of turbulence.

This behavior changes as the transport transitions to the high-turbulence, low-confinement regime ($H_{98y2} < 1.0$, purple points), which corresponds to the avalanche-like state. 
Here, the scaling weakens significantly, flattening to $n_{e,\mathrm{core}} \propto \tilde{n}_\mathrm{rms}^{0.17}$. 
The avalanche-like fluctuations may effectively prevent any significant increases in density, leading to a saturation of the core density profile. 
This saturation occurs at densities far exceeding the Greenwald limit, approaching $2n_\mathrm{G}$.
The weak overall scaling of density with turbulence ($n_\mathrm{core} \propto \tilde{n}_\mathrm{rms}^{0.26}$) is consistent with the weak macroscopic power scaling of the core density limit ($n_{e,\mathrm{core}} \propto P_\mathrm{SOL}^{0.27\pm0.03}$) found previously.

\section{Conclusion and Discussion}
\label{sec:conclusion}

This study re-examined the underlying physics of the density limit using negative triangularity plasmas on the DIII-D tokamak. 
Our findings suggest that the density limit, at least in NT plasmas without an H-mode edge, is not a singular disruptive boundary, but a saturation phenomenon, or a `soft' limit, governed by distinct physical mechanisms: an edge density limit triggered by radiative cooling, and a core density limit driven by avalanche-like turbulent transport.
This study also demonstrates sustained, non-disruptive operations at densities up to 1.8 times the Greenwald limit with substantial auxiliary heating.

At the plasma edge, the density limit manifests as an abrupt saturation event that is linked to the formation of a non-disruptive, radiation front on the high-field side.
While geometrically similar to a disruptive MARFE, this radiative structure acts as a quasi-stable mode that leads to a non-adiabatic, high-turbulence regime and hence may clamp the edge density at a level below the Greenwald limit.

In contrast, the core density is \emph{not} directly constrained by the edge cooling and continues to rise beyond the Greenwald limit. 
Its saturation is instead associated with a fundamental shift in the nature of core turbulence. 
As the density increases, core transport transitions from a diffusive state to one characterized by intermittent, avalanche-like dynamics. 
This is supported by signatures including heavy-tailed PDFs, $1/f$-type power spectra, and elevated Hurst exponents. 
This persistent, non-diffusive transport regime prevents further increases in the core density.

The dual mechanisms of regional density saturation are consistent with the disparate power scalings observed for the core ($n_\text{e,core} \propto P_\mathrm{SOL}^{0.27}$) and edge ($n_\text{e,edge} \propto P_\mathrm{SOL}^{0.42}$). 
The power-dependent nature of this limit and its link to transport physics resembles density limits observed in stellarators. 
This suggests an underlying physics of power balance and transport that transcends specific magnetic configurations, pointing to a more fundamental basis for the density limit in magnetically confined plasmas.

These findings also present a potential strategy for achieving high Greenwald fraction operation in future tokamaks, particularly those utilizing negative triangularity configurations. 
Success hinges on addressing the distinct physical mechanisms governing the density limit in the core and edge. 
The edge radiation front can be mitigated through advanced divertor designs and impurity control. 
Techniques to suppress core turbulence, such as strong rotational shear flows and Shafranov shift, can be developed to optimize density profiles and confinement. 
By using targeted controls for the distinct edge radiative and core turbulent limits, the operational space can be extended more efficiently than by relying on heating power alone.

A further question for the broader applicability of these findings is whether these power scaling and saturation mechanisms manifest in positive triangularity (PT) L-mode plasmas. 
However, conducting a comparable study in PT is challenging in present tokamaks. 
The exploration of a power-dependent density limit requires high auxiliary heating, which typically triggers a transition to H-mode in PT configurations. 
This moves the plasma into a different regime, where the H-mode density limit (HDL) is governed by distinct pedestal physics and not the L-mode mechanisms detailed in this work. 
Consequently, L-mode density limit studies in PT are restricted to low-power regimes that are often disruptive due to radiative condensation instabilities, preventing a clear investigation of the saturation phenomena. 
Further work is also required to establish the causal links between the HFS radiation front and the complex turbulence response.

\section*{Acknowledgments}

The authors gratefully acknowledge the invaluable support and contributions of the entire DIII-D team in performing this experiment.
This material is based upon work supported by the U.S. Department of Energy, Office of Science, Office of Fusion Energy Sciences, using the DIII-D National Fusion Facility, a DOE Office of Science user facility, under Awards Nos.~DE-FC02-04ER54698, DE-SC0019352, DE-SC0019004, DE-FG02-97ER54415, DE-AC52-07NA27344, DE-FG02-08ER54999 and DE-SC0016154.
The author (P.H.D.) acknowledges support from LLNL-led SciDAC ABOUND Project SCW1832.
The author (O.S.) performed this work within the framework of the EUROfusion Consortium, via the Euratom Research and Training Programme (Grant Agreement No.~101052200-- EUROfusion) and funded by the Swiss State Secretariat for Education, Research and Innovation (SERI). Views and opinions expressed are however those of the author(s) only and do not necessarily reflect those of the European Union, the European Commission, or SERI. Neither the European Union nor the European Commission nor SERI can be held responsible for them.

Disclaimer: This report was prepared as an account of work sponsored by an agency of the United States Government. Neither the United States Government nor any agency thereof, nor any of their employees, makes any warranty, express or implied, or assumes any legal liability or responsibility for the accuracy, completeness, or usefulness of any information, apparatus, product, or process disclosed, or represents that its use would not infringe privately owned rights. Reference herein to any specific commercial product, process, or service by trade name, trademark, manufacturer, or otherwise does not necessarily constitute or imply its endorsement, recommendation, or favoring by the United States Government or any agency thereof. The views and opinions of authors expressed herein do not necessarily state or reflect those of the United States Government or any agency thereof.

\bibliography{nt_density_limit}
% \printbibliography

\end{document}